\documentclass[aps,twocolumn,showpacs,superscriptaddress]{revtex4}
\usepackage{amsmath}
\usepackage{amssymb}
\usepackage{epsfig}
\usepackage{color}
\usepackage[colorlinks]{hyperref}
\hypersetup{colorlinks,citecolor=red,linkcolor=blue,urlcolor=blue}
\usepackage{graphicx,amsmath}
\begin{document}

\title{Some features of high-temperature superconductivity on flat
bands}
\author{V. R. Shaginyan}\email{vrshag@thd.pnpi.spb.ru}
\affiliation{Petersburg Nuclear Physics Institute named by B.P.
Konstantinov of National Research Center "Kurchatov Institute",
Gatchina, 188300, Russia}\affiliation{Department of Physics, Clark
Atlanta University, Atlanta, GA 30314, USA}\author{A. Z. Msezane}
\affiliation{Department of Physics, Clark Atlanta University,
Atlanta, GA 30314, USA}
\author{S. A. Artamonov} \affiliation{Petersburg Nuclear Physics
Institute named by B.P. Konstantinov of National Research Center
"Kurchatov Institute", Gatchina, 188300, Russia}

\begin{abstract}
In this letter, we examine how the presence of flat band leads to
the formation of a high-temperature superconductor even in the case
of repulsive pairing interactions. We also show that in the case of
flat bands, the high-temperature superconducting state deforms the
flat band, tilting it and making the effective mass finite. As a
result, neither the superfluid weight nor the supercurrent
disappear. Our results are in good agreement with experimental
data.
\end{abstract}


\maketitle

\section{Introduction}\label{I}

In the standard Bardeen-Cooper-Schrieffer (BCS) theory, attractive
forces such as the electron-phonon interaction form Cooper pairs.
However, in 1965, Kohn and Luttinger demonstrated that even if the
pure interaction between fermions is strictly repulsive (e.g.,
Coulomb repulsion), quantum mechanical corrections and dynamical
screening by the Fermi surface can change the sign of the
interaction in certain angular momentum channels. Thus,
higher-order quantum scattering processes lead to a change in the
repulsive interaction at certain momentum transfers, such as $q
\approx 2p_F$, where $p_F$ is the Fermi momentum. This leads to an
effective attraction, allowing fermions to form pairs, see e.g.
\cite{kag,maiti,dalal,kel}.

Flat bands formed by the topological fermion condensation quantum
phase transition (FCQPT) are fixed at the chemical potential $\mu$,
being shaped by strong electron-electron interactions within the
region $p_f\geq p\geq p_i$ \cite{ks,vol,ksk},
\begin{equation}\label{mu}
\varepsilon({\bf p})-\mu=0; \,\,\,{\rm if}\,\,\, p_f\geq p\geq p_i,
\end{equation}
where $\varepsilon({\bf p})$ is single-particle energy and $\mu$ is
the chemical potential. The occupation numbers of quasiparticles
$n({\bf p})$ become fractional within the flat band range given by
Eq. \eqref{mu}
\begin{equation}\label{mun}
1>n({\bf p})>0.
\end{equation}
These flat bands have influenced the landscape of modern condensed matter physics,
which studies strongly correlated Fermi systems
\cite{ks,ksk,vol,khod97,phys_rep,book20,
Volovik,vol15,prl20,bern,bern_prl,torma,tian,torma1}. For example,
in the case of three-dimensional space, the standard Fermi surface
splits and expands, forming a thickened three-dimensional Fermi
volume \cite{ks,ksk}. This thickened Fermi volume forms the new
state of matter with the universal scaling behavior of the
thermodynamic properties \cite{book20}. Transport properties, that
is, the linear dependence of resistivity $\rho$ on temperature $T$,
$\rho(T)\propto T$, are manifested in heavy-fermion (HF) metals,
high-temperature superconductors, and twisted graphene
\cite{book20,khod94}. Volovik demonstrated that the fermion
condensation is essentially formed by topological phase transition
in momentum space, leading to a new class of Fermi liquids
\cite{vol}. In Landau's Fermi liquid theory, electronic properties
were understood primarily in terms of the kinetic energy of
quasiparticles: How quasiparticles move within a crystal. Flat
bands have changed this paradigm. By reducing the kinetic energy to
zero, flat bands make electron-electron interactions the absolutely
dominant force determining the properties of strongly correlated
Fermi systems.

As a result, the BCS theory \cite{bcs,genn} has also undergone
changes; for example, in the case of high-temperature (high-$T_c$)
flat-band superconductors, the critical temperature $T_c$ is
strictly linear \cite{ks,ksk,khod97}
\begin{equation}\label{TC}
T_c\propto \Delta_1\propto \lambda_0,
\end{equation}
rather than being exponentially suppressed
$T_c\propto\exp{(-1/\lambda_0N(0))}$. From Eq. \eqref{TC} it is
clear that the flat band significantly increases the critical
temperature $T_c$ of high-temperature (high $T_c$) superconductors.
Here $\lambda_0$ is the superconducting coupling constant,
$\Delta_1$ is the maximum value of the superconducting gap and
$N(0)$ is the density of states at the Fermi surface. According to
Eq. \eqref{TC}, flat bands are considered a pathway to
high-temperature superconductivity, see, for example,
\cite{Volovik, torma3}, while recent experimental data may confirm
these long-standing predictions of \cite{minkov}.

In this letter, we show that the critical temperature
$T_c\propto\lambda_0$ in both cases, when $\lambda_0$ corresponds
to an attractive pairing interaction, $\lambda_0<0$, and when
$\lambda_0$ corresponds to a repulsive interaction, $\lambda_0>0$.
We also show that neither the superfluid weight (stiffness) $D_s$
nor the supercurrent vanish in the flat band case, since the flat
band in question is tilted by the superconducting state, making the
effective mass $M^*$ finite.

\section{Superfluid weight in the case of flat bands}

In standard wide-band gap Bardeen-Cooper-Schrieffer (BCS)
superconductors, at zero temperature $T = 0$, all electrons
participate in the formation of the superfluid density. The
superfluid weight is described by the pure limit equation:
\begin{equation}\label{rigid}
D_{s}\simeq \frac{n_ee^2}{M^*},
\end{equation}
where $n_e$ is the electron carrier density, $e$ is the electron
charge, and $M^*$ is the effective mass. As can be seen from Eq.
\eqref{rigid}, heavy electrons suppress the  superfluid weight
$D_s$. If the band narrows, $M^*$ increases, which leads to a
decrease in $D_s$. In strongly interacting, topologically
nontrivial flat bands pinned at the chemical potential $\mu$ (for
example, in twisted bilayer graphene), the energy dispersion
$\varepsilon({\bf p})$ is constant. Therefore, the effective mass
becomes infinite, $M^*\to\infty$, which, according to common sense,
means that quasiparticles cannot move, see e.g. \cite{tian}.
Surprisingly, $D_{s}$ does not drop to zero. To explain this fact,
it is usually assumed that the finite values of both $D_{s}$ and
$T_c$ are completely determined by quantum-geometric origin, see
e.g. \cite{tian,herzog,torma3}. On the other hand, experimental
data show that $T_c$ is proportional to the velocity $V_F$ of
quasiparticles at the Fermi surface, $T_c\propto V_F$,
\cite{mac,epl22,jlett25}. In this case, the disappearance of the
Fermi velocity ($V_F\propto 1/M*\to 0$) corresponds to the presence
of flat band, while $T_c\to0$ and the superconducting state
vanishes, not being restored by the quantum geometry approach.
Thus, at present, an approach based solely on quantum geometry
cannot explain the behavior of flat-band superconductivity.

To clarify this important point, we turn to the traditional
equations of BCS theory, applicable in the case of flat bands
\cite{jlett25}, linking the single-particle energy
$\varepsilon({\bf p})$ with the superconducting gap $\Delta({\bf
p})$ \cite{bcs,genn}
\begin{equation}\label{7*}
\varepsilon({\bf p})-\mu\ =\Delta({\bf p})\frac{1-2v^2({\bf p})}
{2v({\bf p})u({\bf p})}.
\end{equation}
At $T=0$ the coherence factors $v({\bf p})$, and $u({\bf p})$ are
given by
\begin{equation}\label{3*}
n({\bf p})=v^2({\bf p}),\,\, u^2({\bf p})+v^2({\bf p})=1,
\end{equation}
and the order parameter $\kappa({\bf p})$ becomes
\begin{equation}\label{4*}
\kappa({\bf p})=v({\bf p})u({\bf p})=\sqrt{n({\bf p})(1-n({\bf
p}))}.
\end{equation}
Here $n({\bf p})$ is given by Eq. \eqref{mun}.  The superconducting
gap $\Delta({\bf p})$ is given by
\begin{equation}\label{100*}
\Delta({\bf p})=-\frac{1}{2}\int\lambda_0 V({\bf p},{\bf p}_1)
\frac{\Delta({\bf p}_1)}{E({\bf p}_1)}\frac{d{\bf p}_1}{4\pi^2}.
\end{equation}
Here the excitation energy $E({\bf p})$ is defined by the
Bogoliubov quasiparticles
\begin{equation}\label{11*}
E({\bf p})=\sqrt{(\varepsilon({\bf p})-\mu)^2+\Delta^2({\bf p})}.
\end{equation}
Equations \eqref{7*}---\eqref{11*} are the standard ones of the BCS
theory \cite{bcs,genn}, defining the superconducting state with
Bogoliubov quasiparticles and the maximum value of the
superconducting gap $\Delta_1\ \sim 10^{-3} \varepsilon_F$,
provided that the compound under consideration has not undergone
topological FCQPT with the formation of flat bands, when the gap is
given by Eq. \eqref{TC}. Thus, we see that the BCS formalism works,
but the existence of flat bands leads to the appearance of new
features \cite{jlett25} that protect the superconducting state from
destruction by the flat band itself, as follows from Eq.
\eqref{rigid}.

\begin{figure}[!ht]
\begin{center}
\includegraphics[width=0.50\textwidth]{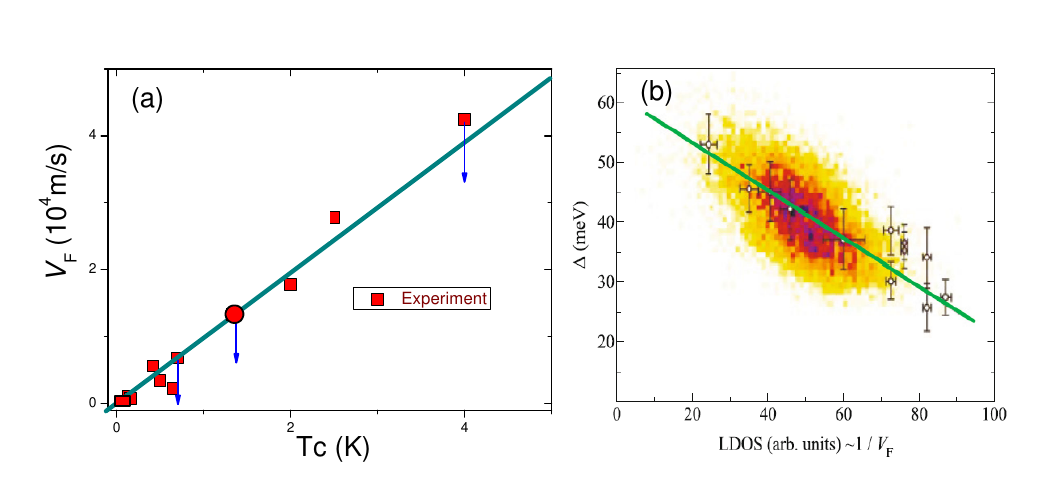}
\end{center}
\vspace*{-0.8cm} \caption{Experimental data illustrating the
validity of Eq. \eqref{15}.\,\, (a) Data (shown as squares) of the
dependence of the average Fermi velocity $V_F$ on the critical
temperature $T_c$ for graphene (MATBG) \cite{mac}. The downward
arrows indicate that $V_F\leq V_0$, where $V_0$ is the maximum
value, indicated by the red square. The theoretical explanation is
represented by the solid straight line. (b) The Figure is adapted
from \cite{nat01} and shows the experimental dependence of the
superconducting gap $\Delta$ on the integrated local density of
states collected on the high-$T_c$ superconductor $\rm
Bi_2Sr_2CaCu_2O_{8+x}$. Here $\rm x$ is the oxygen doping
concentration. Darker colors correspond to more data points with
the same integrated local density of states and the same size of
the gap $\Delta$ \cite{nat01}. The straight blue line shows the
average value of $\Delta$ as a function of the integrated local
density of states (LDOS) that is proportional to
$1/V_F$.}\label{fig05}
\end{figure}
Now let us look at these protective features. From Eq. \eqref{7*}
it is clear that a high$-T_c$ superconducting state operates within
the well-known "self-help" paradigm. Indeed, one takes into account
the "terrible" Eq. \eqref{rigid} and immediately eliminates the
flat band and reduces the value of the effective mass $M^*$ based
on Eq. \eqref{7*}. The effective mass voluntarily becomes
$M^*\propto 1/\Delta_1$ \cite{pla2000,phys_rep,book20,jlett25},
which invalidates the disappearance of the superfluid weight, see
Eq. \eqref{rigid}. One then checks whether the "self-help" paradigm
has done a good job: From Eq. \eqref{7*} it follows that
$\varepsilon({\bf p})-\mu\simeq\Delta$, substituting this relation
into Eq. \eqref{100*}, one arrives at Eq. \eqref{TC}, concluding
that the result is consistent with the "well done" principle.  As a
result, the "dangerous" flat band is cleverly eliminated. Hence Eq.
\eqref{TC} continues to work, and the supercurrent persists.

From Eq. \eqref{7*} it follows that if $\Delta\to 0$, the flat band
is then restored, resulting in $T_c\to 0$. Further consideration
leads to the result that contradicts the BCS theory
\cite{mac,epl22,jlett25}, and follows from Eq. \eqref{7*}
\begin{equation}\label{15}
N(0)\propto M^*_{FC}\propto 1/\Delta_1\propto 1/T_c\propto 1/V_F,
\end{equation}
where $V_F\propto p_F/M^*_{FC}$ is the Fermi velocity, see Fig.
\ref{fig05}, (a) and (b).

Measurements of $V_F$ as a function of $T_c$ \cite{mac} are shown
in Fig. \ref{fig05} (a). The data agree well with Eq. \eqref{15}
\cite{pla2000,phys_rep,jlett25}.  An unusual behavior of
$\Delta\propto V_F$, corresponding to that observed on graphene
\cite{mac}, as seen from Fig.  \ref{fig05} (b), was obtained in
measurements on the high$-T_c$ superconductor $\rm
Bi_2Sr_2CaCu2O_{8+x}$ \cite{epl22,nat01}. Thus, $V_F\to 0$ and also
$T_c\to 0$, as can be seen from Figs. \ref{fig05}(a) and (b). This
result demonstrates that the flat band is tilted by superconducting
state, as shown in Figs. \ref{fig05} (a) and (b), and seen from Eq.
\eqref{15}. It should be noted that it is clear from the Figs. 
\ref{fig05} (a) and (b) that the maximum critical temperatures
$T_c$ do not correspond to the minima of the Fermi velocity $V_F$,
as would be the case in any BCS-type theory \cite{mac}.

It is worth noting that from Eq. \eqref{15} it directly follows
that HF metals, which have extremely high values of the effective
mass $M^*$, have extremely low $T_c$ \cite{pag1,khod15,SHVM}. For
example, the maximum value of $T_c$ is achieved in the HF metal
$\rm CeCoIn_5$ at $T_c\simeq 2.3$ K. While $M^*$ has a small value
compared to other corresponding HF metals \cite{pag1,SHVM}. This
observation is in good agreement with the experimental data
presented in Fig. \ref{fig05}. Thus, it is obvious that Eq.
\eqref{15} is in good agreement with experimental data obtained
from measurements on HF metals, which are unconventional
superconductors, as well as on high$-T_c$ superconductors, which
are also unconventional superconductors. As a result, we
confidently conclude that the superconducting state operates
effectively within the well-known "self-help" paradigm in
accordance with the fermion condensation theory
\cite{ks,ksk,khod97,book20}.

\section{Superconductivity with repulsive interaction}

Here we begin with a thermodynamic consideration of
superconductivity in the presence of flat bands formed by fermion
condensate (FC). Due to the strong repulsive interaction, the
topological FCQPT arises in the system, and FC with a flat band
state emerges. At $T=0$ the quasiparticle distribution $n({\bf p})$
represented by the step function becomes unstable and fragmented.
In the flat band region, $n({\bf p})$ takes fractional values $0<
n({\bf p})< 1$ even at $T\to0$, as can be seen from Eq. \eqref{mun}
\cite{ks}. Thus, the FC state creates an anomalously elevated
density of states and macroscopic degeneracy of single-particle
energy states. When calculated, this FC state results in a residual
entropy $S_{0}$ that remains finite even as $T$ approaches absolute
zero: $S(T\to0)=S_0>0$ \cite{khod94} . This violates the third law
of thermodynamics, or Nernst's theorem. To resolve this physical
contradiction, highly degenerate flat-band states must acquire at
least a tiny but finite dispersion, or energy gap, at $T\to 0$.
Thus, the FC state loses its residual entropy $S_0$ at zero
temperature, undergoing phase transitions, to satisfy Nernst's law.
As a result, the FC state serves as a stimulator for competing
phase transitions at low temperatures. In materials such as HF
metals or high-$T_c$ superconductors, this leads to a complex phase
diagram, where the competition between these phases determines the
behavior of the strange metal, characterized by non-Fermi liquid
(NFL) behavior \cite{phys_rep}. Thus, at $T\to 0$ the system under
consideration must undergo a phase transition or a sequence of
phase transitions, and the predominant phase transition must be a
superconducting phase transition, since the FC state is
characterized by the superconducting order parameter, see Eq.
\eqref{4*}.

Indeed, the fact that $\kappa({\bf p})\neq 0$, see Eq. \eqref{4*},
even for $\lambda_0=0$ (and hence $\Delta_1=0$), shows that the BCS
pairing is initiated by the strong repulsive interaction of
quasiparticles and wins the competition between phase transitions
even at $T\to0$, as exemplifies by the canonical HF metal $\rm
YbRh_2Si_2$ \cite{steg}. Pairing with such unusual properties occur
due to the flat band formed as a result of the fermion
condensation. Moreover, from Eqs. \eqref{TC} and \eqref{100*} it is
clear that a non-trivial solution $\Delta_1\neq0$ can exist even if
$\lambda_0$ becomes negative, i.e. repulsive \cite{khod97}. This
result completely contradicts the BCS theory. Indeed, according to
Eqs. \eqref{TC} and \eqref{100*}, $\lambda_0$ is nothing more than
a proportionality factor in the relationship between the gap
$\Delta_1$ and the anomalous density $\kappa({\bf p})$. In contrast
to the BCS theory, $\kappa({\bf p})$ is practically independent of
the interaction $\lambda_0$ (see Eq. \eqref{4*}). It is formed by a
strong repulsive interaction, which produces a thickened Fermi
volume distribution $n({\bf p})$ given by Eq. \eqref{mun}. Thus, we
come to the conclusion that high$-T_c$ superconductors with
repulsive pair interaction must be based on flat-band compounds,
since only in this case high$-T_c$ superconductivity is ensured by
Eq. \eqref{TC}. Experimental data show that the corresponding flat
bands in twisted graphite are formed by electron correlations
\cite{ghosh}, while the long-awaited room-temperature
superconductivity with flat bands, see e.g. \cite{Volovik, torma3},
is observed in graphite intercalated with calcium and ammonia
solutions \cite{minkov}. We note that in the case of effective
attraction, allowing fermions to form pairs, $T_c$ is determined by
the BCS Eq. \eqref{100*}, which allows only an exponentially small
value of $T_c$ in the absence of the corresponding flat band, see
e.g. \cite{kag,maiti,dalal,kel}.

\section{Conclusions}\label{conc}

We have examined how the presence of flat band leads to the
formation of  high$-T_c$ superconductors even in the case of
repulsive pairing interactions. In the framework of the FC theory,
we have also shown that in the case of flat bands the superfluid
weight does not vanish, which determines the existence of a
supercurrent. For the first time, we have shown how the flat band
of high$-T_c$ superconductors changes under the influence of the
superconducting state, tilting in such a way as to avoid the
complete loss of both superfluid weight and the supercurrent. Our
studies of important experimental data convincingly demonstrate
that the topological FCQPT is an inherent feature of many strongly
correlated Fermi systems and can be considered a universal cause of
their non-Fermi-liquid properties, which are not observed in metals
described by the Landau Fermi-liquid theory. Moreover, the FC
theory can explain the complex behavior of strongly correlated
Fermi systems represented by high$-T_c$ superconductors.

\section{funding} This work was supported by ongoing
Petersburg Nuclear Physics Institute named by B.P. Konstantinov of
National Research Center "Kurchatov Institute" funding. No
additional grants to carry out or direct this particular research
were obtained.

\section{conflicts of interest}{The authors of this work
declare that they have no conflicts of interest.}


\begin{thebibliography}{99}

\bibitem{kag} M.Y. Kagan, D.V. Efremov, M.S. Mar$^,$enko, and
V.V. Val$^,$kov,  J. Supercond. Nov. Magn. {\bf  26}, 2809 (2013).

\bibitem{maiti}  S. Maiti and  A.V. Chubukov, AIP Conf. Proc. {\bf 1550}, 3 (2013).

\bibitem{dalal} A. Dalal, J. Ruhman, and V. Kozii,
Sci. Post. Phys. Core {\bf 9}, 034 (2026).

\bibitem{kel} A. Keles, X. Li, and E. Zhao, Phys. Rev. B {\bf 109}, 054519 (2024).

\bibitem{ks} V.A. Khodel and V.R. Shaginyan,
JETP Lett. {\bf 51}, 553 (1990).

\bibitem{vol} G.E. Volovik, JETP Lett. {\bf 53}, 222 (1991).

\bibitem{ksk} V.A. Khodel, V.R. Shaginyan, and V.V. Khodel, Phys. Rep.
{\bf 249}, 1 (1994).

\bibitem{khod97} J. Dukelsky, V.A. Khodel, P. Schuck, and
V.R. Shaginyan, Z. Phys. {\bf 102}, 245 (1997).

\bibitem{phys_rep} V.R. Shaginyan, M.Ya. Amusia, A.Z. Msezane,
and K.G. Popov, Phys. Rep. {\bf 492}, 31 (2010)

\bibitem{book20} M.Ya. Amusia and V.R. Shaginyan,
{\it Strongly Correlated Fermi Systems: A New State of Matter},
Springer Tracts in Modern Physics Vol. {\bf 283} (Springer Nature
Switzerland AG, Cham, 2020).

\bibitem{torma}  R.P.S. Penttil\"a, K.-E. Huhtinen, and P. T\"orm\"a,
Commun. Phys. {\bf 8}, 50 (2025).

\bibitem{bern} P. T\"orm\"a,  S. Peotta, and B.A. Bernevig,
Nat. Rev. Phys. {\bf 4}, 528 (2022).

\bibitem{torma1}  G. Jiang, P. T\"orm\"a, and Y. Barlas,
PNAS  {\bf 122},  e2416726122 (2025).

\bibitem{tian} H. Tian, X. Gao, Y. Zhang, S. Che, T. Xu, P. Cheung,
K. Watanabe, T. Taniguchi, M. Randeria, F. Zhang, C.N. Lau, and M.
W. Bockrath, Nature {\bf 614}, 440 (2023).

\bibitem{bern_prl} V. Peri, Z.D. Song,
B.A. Bernevig, and S.D. Huber, Phys. Rev. Lett. {\bf 126}, 027002
(2021).

\bibitem{Volovik} T.T. Heikkila and G.E. Volovik, Flat bands as a route
to high-temperature superconductivity in graphite. Springer Series
in Materials Science, Vol. {\bf 244} (Springer Nature Switzerland
AG, Cham, 2016)

\bibitem{vol15} G.E. Volovik, Phys. Scr. T{\bf 164}, 014014 (2015).

\bibitem{prl20} P. Rosenzweig, H.
Karakachian, D. Marchenko, K. K\"uster, and U. Starke, Phys. Rev.
Lett. {\bf 125}, 176403 (2020).

\bibitem{khod94} M.V. Zverev, V.A. Khodel, and V.R. Shaginyan,  JETP
Lett. {\bf 60}, 541 (1994).

\bibitem{bcs} J. Bardeen, L.N. Cooper, and J.R. Schriffer,
Phys. Rev. {\bf 108}, 1175 (1957).

\bibitem{genn} P.G. De Gennes, {\it
Superconductivity of Metals and Alloys}, (W.A. Benjamin Inc. New
York, Amsterdam, 1966).

\bibitem{torma3} R.P.S. Penttil\"'a, K-E. Huhtinen, and P.
T\"orm\"a,  Commun. Phys. {\bf 8}, 50 (2025).

\bibitem{minkov} V.S. Minkov, V. Ksenofontov, and M.I. Eremets,
Carbon Trends, {\bf 24}, 100672 (2026).

\bibitem{herzog} J. Herzog-Arbeitman, V. Peri, F. Schindler, S.D. Huber, and B.A.
Bernevig, Phys. Rev. Lett. 128, 087002 (2022).

\bibitem{mac} W. Qin, B. Zou, and A.H. MacDonald,
Phys. Rev. B {\bf 107}, 024509 (2023).

\bibitem{epl22} V.R. Shaginyan, A.Z. Msezane, M.Ya. Amusia,
and G.S. Japaridze, EPL {\bf 138}, 16004 (2022).

\bibitem{jlett25} V.R. Shaginyan, A.Z. Msezane, and S.A. Artamonov,
JETP Lett. {\bf 122}, 158 (2025).

\bibitem{pla2000} M.Ya. Amusia and V.R. Shaginyan,
Phys.  Lett. A{\bf 275}, 124 (2000).

\bibitem{nat01} S.H. Pan, J.P. O'Neal, R.L. Badzey, C. Chamon,
H. Ding, J.R. Engelbrecht, Z. Wang, H. Eisaki, S. Uchida, A. K.
Gupta, K.W. Ng, E.W. Hudson,  K.M. Lang, and J.C. Davis, Nature
{\bf 413}, 282 (2001).

\bibitem{pag1} J. Paglione, M. A. Tanatar, D. G. Hawthorn,
E. Boaknin, R. W. Hill, F. Ronning, M. Sutherland, L. Taillefer, C.
Petrovic, and P. C. Canfield,  Phys. Rev. Lett. {\bf 91}, 246405
(2003).

\bibitem{khod15} V.A. Khodel, J.W. Clark, K.G. Popov, and V.R. Shaginyan,
JETP Lett. {\bf 101}, 413 (2015).

\bibitem{SHVM} S. K. Shrivastava, IJEAT {\bf 9}, 1378 (2020).

\bibitem{steg} E. Schuberth, M. Tippmann, L. Steinke, S. Lausberg,
A. Steppke, M. Brando, C. Krellner, C. Geibel, R. Yu, Q. Si, and F.
Steglich, Science {\bf 351}, 485 (2016).

\bibitem{ghosh} A. Ghosh, S. Chakraborty, R. Dutta, A. Agarwala,
K. Watanabe, T. Taniguchi, S. Banerjee, N. Trivedi, S. Mukerjee,
and A. Das, Nat. Phys. {\bf 21}, 732 (2025).

\end{thebibliography}
\end{document}